\documentclass{ws-tpe}

\usepackage[super]{cite}
\usepackage{xcolor}
\usepackage{graphicx}
\usepackage{stfloats}
\usepackage{enumitem}
\usepackage[verbose,hypertexnames=false]{hyperref}
\hypersetup{colorlinks=false,allbordercolors=blue,pdfborderstyle={/S/U/W 1}}
\usepackage{multicol}

  \newcommand{\corr}[1]{#1}  

  \newcommand{\rev}[1]{#1}  

  \newcommand{\revIII}[1]{#1}  

\begin{document}

\catchline{}{}{2025}{}{}
\markboth{H. Maus et al.}{Developing an LLM-Based Feedback System Grounded in Evidence-Centered Design to Support Physics Problem Solving}

\title{\corr{Developing an LLM-Based Feedback System Grounded in Evidence-Centered Design to Support Physics Problem Solving
\\}}

\author{Holger Maus$^{\ast}$, Fabian Kieser$^{\dagger}$, Stefan Petersen$^{\ast}$,  Peter Wulff$^{\ddagger}$, and Paul Tschisgale$^{\ast}$}

\address{$^{\ast}$Department of Physics Education, Leibniz Institute for Science and Mathematics Education\\ Kiel, Germany}

\address{$^{\dagger}$Department of Physics Education Research, Free University of Berlin\\ Berlin, Germany}

\address{$^{\ddagger}$Department of Physics and Physics Education Research, Ludwigsburg University of Education\\ Ludwigsburg, Germany}











\maketitle

\begin{history}
\received{Day Month Year}
\revised{Day Month Year}
\accepted{Day Month Year}
\end{history}

\begin{abstract}
Generative AI offers new opportunities for individualized and adaptive learning, e.g., through large language model (LLM)-based feedback systems. While LLMs can produce factually correct feedback for relatively straightforward conceptual tasks, delivering high-quality feedback for tasks that require advanced domain expertise—such as physics problem solving—remains a substantial challenge.
This study presents the design and implementation of an LLM-based feedback system for physics problem solving grounded in evidence-centered design and reports a first evaluation within the German Physics Olympiad. \revIII{Participants rated the usefulness and correctness of the generated feedback for each implemented problem. The collected ratings indicate that the feedback was generally perceived as useful and highly correct.} However, an in-depth analysis revealed that the feedback contained errors in 20\% of cases—errors that often went unnoticed by the students.
We discuss the risks associated with uncritical reliance on LLM-based feedback and outline potential directions for generating more adaptive and reliable LLM-based feedback in the future.
\end{abstract}


\begin{multicols}{2}
\section{Introduction}
Recent advances in artificial intelligence (AI), particularly in large language models (LLM), have opened promising opportunities to provide automated, individualized, and meaningful LLM-generated feedback in a range of disciplines, including physics \cite{yin_using_2024, chen_grading_2025, cheng_science_2025}. To date, however, most LLM-based feedback systems—and the corresponding research—have focused primarily on advancing students' factual knowledge and conceptual understanding. 
\corr{It still remains largely unclear to what extent such systems can also provide useful and factually correct feedback on more complex and multifaceted activities, such as problem solving, which is widely recognized as a key 21st-century skill and considered particularly important for individuals pursuing physics and other science-related careers \cite{bransford_how_2000, frey_teaching_2022}.}

To become a proficient problem solver, continuous and targeted deliberate practice is essential \cite{Ericsson.1998}. In particular, formative feedback has been shown to play a crucial role in developing students’ physics problem-solving abilities \cite{gaigher_exploring_2007}. However, providing high-quality feedback requires an accurate assessment of students’ problem-solving abilities. 
\corr{This poses a considerable challenge because problem solving is inherently complex. It requires the integrative use of multiple types of knowledge and skills, including sophisticated problem-solving strategies that distinguish experts from novices \cite{burkholder_characterizing_2020}. These include, for instance, identifying the relevant physics concepts, making appropriate idealizations, and applying the necessary mathematical procedures.}

Evidence for these knowledge types and skills needs to be identified in students’ written problem solutions, interpreted in light of the specific problem at hand, and then appropriately addressed in the feedback. Designing a feedback system that performs these tasks automatically in a valid and reliable manner therefore requires substantial domain expertise and assessment expertise.

Evidence-Centered Design (ECD) \cite{mislevy_brief_2003} offers a promising framework for tackling this challenge. By systematically linking the types of knowledge and skills involved in problem solving with the respective evidence observable in students’ problem solutions, ECD provides a structure to assess students' problem solutions and guide feedback generation. In the context of LLM-generated feedback, ECD can serve as a guiding framework to constrain and direct the LLM. A known challenge of LLMs, partly rooted in their training processes, is their tendency to produce generic or “averaged” responses that are insufficiently grounded in the specifics of an expert solution, or to confabulate information entirely \cite{Summerfield.2025}. 
Hence, rather than producing surface-level or holistic feedback, LLMs can be prompted using ECD to identify specific forms of evidence, such as relevant concepts, missing assumptions or reasoning steps, and wrong formulas. This evidence can then be mapped to targeted feedback aligned with the intended types of knowledge and skills.

In this study, we report on the design, implementation, and initial evaluation of an ECD-grounded LLM-based automated feedback system that aims to support students in developing their physics problem-solving abilities. The system is designed to automatically assess students’ physics problem-solving processes and provide individualized analytical, rather than holistic, feedback. We evaluate the system in the context of the German Physics Olympiad, focusing on how participating students perceived the usefulness and correctness of the LLM-generated feedback.

\section{Theoretical Background}
\subsection{Physics Problem Solving}
Problem solving in physics requires the structured and goal-oriented application of physics knowledge and skills to successfully solve problems, i.e., effectively transform an initial state into a desired goal state \cite{smith_toward_1988}. 
 
Research on problem solving across a range of scientific disciplines has identified distinct phases in the problem-solving process \cite{Witte1972PhaseTheorem}. Specifically, problem solving in physics has been shown to involve the application of multiple types of knowledge and skills \cite{friege_types_2006,tschisgaleExploringSequentialStructure2025}. 

Fig.~\ref{fig_problem_solving_knowledge} illustrates physics problem solving as a sequential process comprising prototypical phases, each of which involves the application of specific knowledge types and skills \cite{leonard_using_1996}. While conceptual tasks primarily draw on conceptual knowledge (and sometimes conditional knowledge), genuine problem solving additionally demands procedural knowledge (applying operators), factual knowledge (e.g., explicit formulas for axioms or laws), mathematics skills (since problems in physics often hinge on mathematics \corr{\cite{BingRedish2009,Tegmark.2008}}), and metacognitive knowledge to plan actions, monitor them, and regulate the overall problem-solving process.

\vspace*{0.7cm}
\begin{figurehere}
\centering
\includegraphics[width=3.4 in]{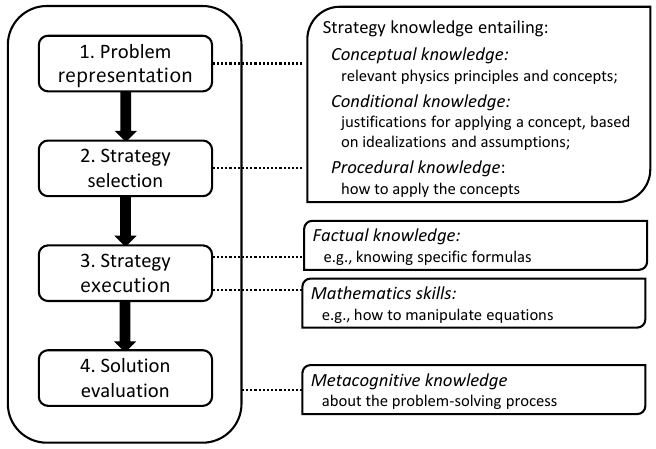}
\caption{Idealized phases of the physics problem-solving process of experts (left) and involved knowledge types and skills essential for successful problem solving (right). \textit{Note:} In addition to metacognitive knowledge, the solution evaluation phase generally involves also all of the other indicated knowledge types and skills.}
\label{fig_problem_solving_knowledge}
\end{figurehere}

Physics problem-solving abilities are typically fostered through well-defined end-of-textbook problems. However, developing expertise in physics problem solving requires continuous and targeted deliberate practice \cite{Kim.2002,Ericsson.1998}, in which feedback and guidance from others, such as domain experts, play a crucial role. 
Feedback is generally most effective when it is both adaptive to the learner’s needs and provided in a timely manner \cite{hattie_power_2007}. However, the provision of adaptive and timely feedback is prohibitive in many real-world settings given resource limitations. Feedback by instructors also tends to be more holistic, rather than analytic, which is less helpful for learners, as it rarely differentiates between the various knowledge types and skills involved in physics problem solving \cite{docktor_assessing_2016, jescovitch_comparison_2021, poldner_assessing_2014}.

\subsection{LLM-Based Automated Feedback}


While in traditional machine learning, building an automated feedback system required large amounts of rather specific training examples, this paradigm changed with the advent of LLMs. LLM-based feedback guided by careful prompting does not require data-intensive or time-consuming model training or fine-tuning. Instead, the primary challenge lies in crafting effective prompts (also referred to as prompt engineering) that ensure the desired performance and output \cite{federiakinPromptEngineeringNew2024}. Another advantage is that LLMs can flexibly address a wide range of student responses in fluent, natural language. \corr{A growing number of research groups have developed or are developing such LLM-based feedback systems \cite{avila_using_2024, xavier_empowering_2025, tufino_notebooklm_2025}}. More importantly, increasing evidence suggests that such LLM-based systems can support learning \cite{wanExploringGenerativeAI2024, kestinAITutoringOutperforms2025} and positively influence academic performance, although effect differs across disciplines \cite{dongExaminingEffectArtificial2025}. In their meta-analysis, Dong et al. found strong effects in nursing, geography, and language learning, but weaker effects in mathematics; however, no conclusions were drawn regarding physics.  Importantly, Dong et al. call for more research into the conditions under which AI is most effective. This call is particularly relevant for physics, as existing evidence has largely focused on conceptual physics tasks rather than genuine physics problem solving \cite{wanExploringGenerativeAI2024}.
Because physics problem solving is more difficult to assess than conceptual understanding and requires feedback on multiple knowledge types and skills, both the effectiveness of LLM-based feedback and the design conditions under which it might be effective remain largely unexamined.

At the same time, a substantial body of research highlights potential risks of using AI in education. Unproductive uses of AI may hinder learning; for instance, when users outsource thinking to the LLM—often referred to as “metacognitive laziness” \cite{fan_beware_2025}, “cognitive debt” \cite{kosmyna_your_2025}, \corr{or "deskilling" \cite{Rafner.2021}}—this may reinforce further dependence on such systems. Particularly, students who heavily rely on LLMs have been found to perform worse in the long term compared to peers who did not use such tools \cite{kosmyna_your_2025,Bastani.2025}. Moreover, on the part of the machine, LLMs are trained to satisfy users, called ``sycophancy'' (insincere flatterers), and may hallucinate/confabulate information \cite{cheng_sycophantic_2026,Summerfield.2025}. LLMs may also produce complete solutions rather than feedback that meaningfully supports learning \cite{Shuster2021_RAG_Hallucination}.
Another issue concerns how students utilize LLM-generated content. While errors in simple calculation tasks or in domains where students already possess strong prior knowledge are often recognized, errors in more complex and difficult settings are often adopted uncritically, a phenomenon described in the literature as ``unreflected acceptance'' \cite{krupp_unreflected_2024, helal_when_2024}. Interestingly, users tend to trust LLM-generated responses more when they include longer explanations, even if the responses are incorrect. \cite{Steyvers2025WhatLLMsKnow}.

Consequently, without careful oversight of both students’ interactions with the feedback system and the technical and pedagogical implementation of the system, such interactions may be detrimental to students.

Many current LLM-based feedback systems provide feedback to students' responses to questions of a predominantly conceptual nature \cite{avila_using_2024}. They rarely take the additional step of engaging students with actual physics problems that require representing a problem from a science perspective, multi-step reasoning, and applying advanced mathematical operations—i.e., genuine physics problem solving. One reason for this gap is that physics problem solving is inherently more complex to assess, and accurate assessment is a prerequisite for providing feedback that is both factually and pedagogically sound. 

In fact, recent advancements in LLM research indicate that LLMs can accurately solve problems in a variety of disciplines, e.g., law, medicine, and even science \cite{OpenAI2023GPT4TechReport}. Even more, recent LLMs seem to approximate expert problem-solving performance in physics \cite{kortemeyer_boiling-frog_2026,tschisgale_evaluating_2025, yuHiPhOHowFar2025}. There is also growing evidence that LLMs can effectively be applied to grading and assessment tasks \cite{mokUsingAILarge2024, kortemeyerGradingAssistanceHandwritten2024a, chen_grading_2025}.
At the same time, LLM-generated responses have been shown to reproduce typical student misconceptions \cite{kieserUsingLargeLanguage2024} and errors \cite{kortemeyer_could_2023}. To mitigate the risk of erroneous model output—particularly in the context of feedback—research suggests using prompt engineering strategies that provide the LLM with model solutions and principles of effective feedback  \cite{sirnoorkarFeedbackThatClicks2025}. 

\subsection{Design-Principles for LLM-Based Feedback Systems}

\corr{In sum, current LLM-based feedback systems demonstrate considerable promise. However, there remains a pressing need for approaches capable of providing reliable and pedagogically sound feedback for complex activities such as students’ physics problem solving.}

\corr{Because the effectiveness of AI appears to be condition-dependent rather than guaranteed \cite{dongExaminingEffectArtificial2025}, the way such a system is designed becomes decisive. We therefore derive the following principles, each responding to a specific concern identified above:}

\begin{itemize}[leftmargin=*, labelsep=2mm]
    \item Feedback for physics problem solving can be automated with the use of recent generative AI tools such as LLMs and implemented online for ease of access\corr{, addressing the resource limitations that make adaptive and timely feedback prohibitive in practice.}
    \item Due to the inherent limitations of even advanced LLMs (particularly hallucinations/confabulations), feedback generation should be grounded in expert solutions while remaining closely tied to learners’ own solution approaches.
    \item Given the tendency of students to accept LLM-generated content in a rather unreflected manner, LLM-based feedback systems should omit provision of complete solutions and entail means for repeated interaction with the LLM on the part of the students.
    \item To iteratively improve the feedback system and remain attentive to students’ needs while keeping pace with developments in AI, empirical testing in real-world settings should be conducted\corr{, directly taking up Dong et al.'s \cite{dongExaminingEffectArtificial2025} call for research into the conditions under which AI is most effective.}
\end{itemize}

\corr{\subsection{Research Questions}}
We describe the development of an LLM-based feedback system for physics problem solving according to the outlined design principles. The system was developed in the context of the German Physics Olympiad, where students are required to demonstrate a solid understanding of physics and strong problem-solving abilities to be successful\cite{tschisgale2024science}. This context provides a suitable setting for evaluating the system, as participants are generally highly interested in physics but heterogeneous in terms of their physics problem-solving abilities \cite{tschisgale_towards_2024}. Accordingly, students can be expected to vary substantially in their problem-solving approaches, in the specificity and sophistication of their physics reasoning, and in the kinds of responses they provide \cite{PhysRevPhysEducRes.21.010111}. 

To evaluate the system, it was made available to participants of the German Physics Olympiad as a voluntary problem-solving training tool. We specifically aim to answer the following research questions (RQ):

\begin{itemize}[leftmargin=*, labelsep=2mm]
\item[1)] \revIII{To what extent is the LLM-generated feedback perceived as useful?}
\item[2)] \revIII{To what extent is the LLM-generated feedback perceived as correct, and to what extent does this perceived correctness align with the actual correctness of the generated feedback?}
\end{itemize}

\section{Design of an LLM-Based Feedback System for Physics Problem Solving}

\subsection{Evidence-Centered Design for Physics Problem Solving}
\label{sec:ecd}
Evidence-Centered Design (ECD) \cite{mislevy_brief_2003} provides a viable framework for LLM-based feedback systems, as it anchors feedback in evidence derived from student responses while simultaneously accounting for the complexities of physics problem solving in a valid and interpretable manner.
More precisely, ECD operates between three interconnected \textit{spaces}, also illustrated in the upper part of Fig.~\ref{fig_ECD_LLM_feedback}:

In the \textit{claim space}, the construct of interest (here: physics problem-solving ability) is first specified. This typically involves decomposing the construct into its constituent knowledge types and skills that students ought to have (or develop), as exemplified for physics problem-solving ability in Fig.~\ref{fig_problem_solving_knowledge}.
Building on this specification, one must then determine what constitutes valid evidence for the identified knowledge types and skills. This is formalized in the \textit{evidence space} through evidence statements. For physics problem solving, these evidence statements are provided in Table~\ref{tab:evidence_statements}.
Subsequently, in the \textit{task space}, tasks are designed to elicit the evidence described in the evidence statements for the targeted knowledge types and skills. In our case, nearly all quantitative and well-defined, multi-step physics problems are suitable, provided that inputting solutions via keyboard is reasonably feasible.
Returning to the \textit{evidence space}, the analysis and interpretation of student-produced evidence require the specification of an evidentiary scheme for each problem. These schemes can be regarded as problem-specific evidence statements that define observable indicators in student responses that demonstrate mastery of the targeted knowledge types and skills. Through these evidentiary schemes, it becomes possible to draw valid inferences regarding the extent to which the knowledge and skills underlying physics problem-solving abilities are present.
\vspace*{13pt}

\begin{figurehere}
\centering
\includegraphics[width=3.5 in]{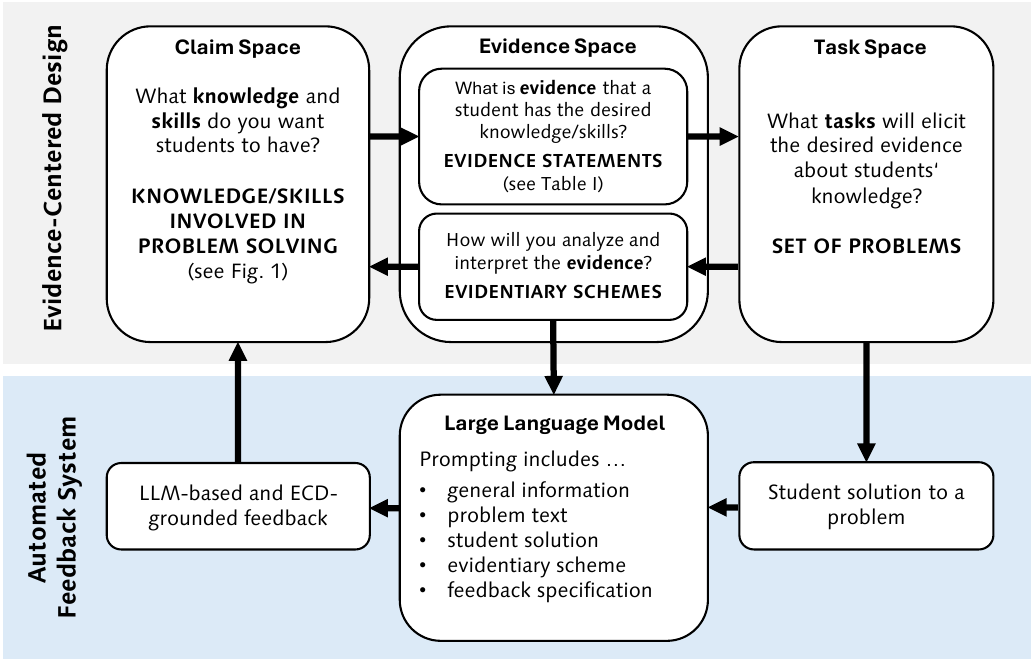}
\caption{Simplified representation of evidence-centered design (adapted from Kubsch et al. \cite{kubsch_toward_2022}) and its integration into the automated LLM-based feedback system.}
\label{fig_ECD_LLM_feedback}
\end{figurehere}

\begin{tablehere}
\tbl{Evidence Statements for Knowledge Types and Skills}
{\small
\begin{tabular}{@{}p{0.8in}p{2.3in}@{}}
\toprule
Knowledge types and skills & Evidence statements \\ \colrule

Conceptual knowledge & Students mention the relevant physics concepts (or principles) \\

Conditional knowledge & Students mention relevant assumptions and idealizations, and how they relate to the application of physics concepts \\

Procedural knowledge & Students describe how the physics concepts would be applied or explicitly apply them \\

Factual knowledge & Students specify physics concepts and relationships between quantities using key formulas or verbal descriptions \\

Mathematics skills & Students correctly apply formula-based or other mathematical procedures \\

Metacognitive knowledge\corr{\footnotemark{}}  & Students' problem solving aligns with the typical sequence of expert problem solving (see Fig.~\ref{fig_problem_solving_knowledge}) \\

\botrule
\end{tabular}}
\label{tab:evidence_statements}
\end{tablehere}
\footnotetext{\corr{Metacognitive knowledge itself—such as planning, monitoring, and regulating the problem-solving process—is not directly observable in a written solution. We therefore use structural alignment with the expert problem-solving sequence as an observable indicator of metacognitive knowledge. This particularly differs from procedural knowledge, which refers to the application of specific operations within individual problem-solving phases.}}


\subsection{Automated Feedback Generation Using LLMs}
\subsubsection{Web Interface}
\begin{figure*}[!t]
\centering
\includegraphics[width=6 in]{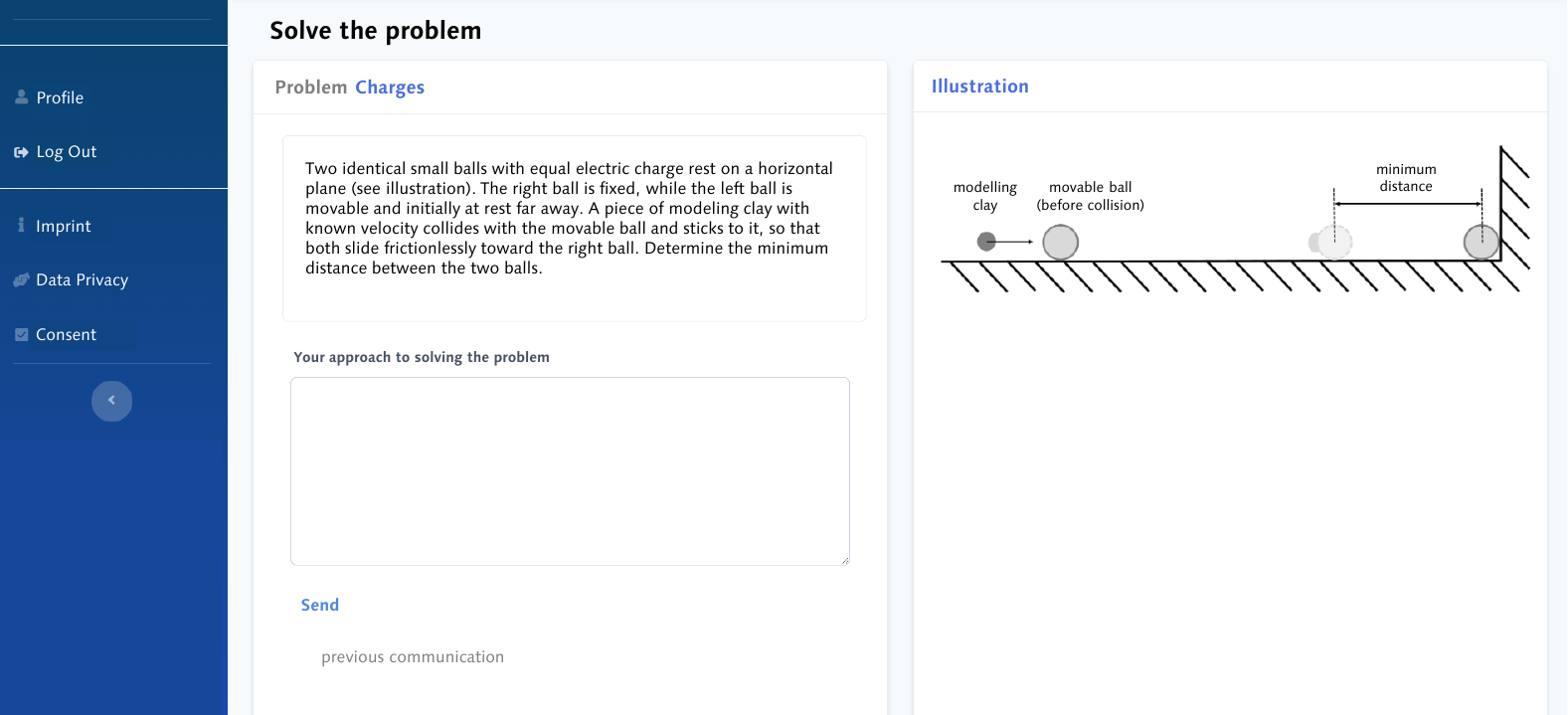}
\caption{Screenshot showing the feedback system's web interface displaying one of the integrated physics problems.}
\label{fig_frontend}
\end{figure*}

The design of the web application for the feedback system is shown in Fig.\ref{fig_frontend}. 
\corr{Students could work on up to six different problems that were implemented in a fixed sequential order.} For each problem, a description of the problem and an accompanying illustration are provided. To enable genuine interaction, the system implements a two-step feedback process: First, students receive LLM-generated feedback on their initial solution approach and are prompted by the feedback system to revise their approach accordingly. Second, a final LLM-generated feedback is provided to the revision. After this, students may proceed to the next problem. Editing can be paused and resumed at any time, and a dashboard allows students to track how many problems they have already completed.

\subsubsection{Backend Prompting}
By providing an LLM with a physics problem, a corresponding student-generated solution, and suitable prompting, it is possible to generate feedback on the solution (see lower part of Fig.~\ref{fig_ECD_LLM_feedback}). To reduce the risk of confabulations, such as physically inaccurate feedback, we included problem-specific evidentiary schemes derived from the ECD approach in the prompt, thereby grounding the LLM-generated feedback in predefined evidentiary schemes. This approach combines the LLM’s capability to produce fluent responses with pedagogically meaningful feedback grounded in the evidence contained in students’ problem-solving approaches.

We used OpenAI’s \textit{GPT-4o} model (snapshot \texttt{gpt-4o-2024-08-06}; hereafter, \textit{GPT-4o}) via the OpenAI API (temperature~=~0.7, top\_p~=~1.0, top\_k~=~0, max\_tokens~=~512), particularly since
\textit{GPT-4o} has been shown to demonstrate advanced apparent physics understanding and problem-solving capabilities \cite{kortemeyer_multilingual_2025, tschisgale_evaluating_2025}.

As already illustrated in the lower part of Fig.~\ref{fig_ECD_LLM_feedback}, the entire prompt processed by \textit{GPT-4o} consists of five components: \textit{general information}, \textit{problem text}, \textit{student solution}, \textit{evidentiary scheme}, and \textit{feedback specification}. While the evidentiary scheme is problem-specific (according to the statements in Table~\ref{tab:evidence_statements}), the general information and the feedback specification each remained almost\footnote{There were minor prompting differences between the first and second feedback round.} identical across all problems.

For clarity, we provide a detailed description of the prompting\footnote{Prompts translated from German to English by the authors.} for a specific physics problem implemented in the feedback system. The prompting starts with \textit{general information}:
\begin{quote}
\small
"You are a helpful tutor for students participating in a physics competition. In this competition, the participants are supposed to learn how to solve physics problems. You are given a student’s approach to a physics problem and should provide short feedback on it." [...]
\end{quote}
The second component is the actual \textit{problem text}. For this example, consider the \textit{Charges} problem which was implemented in the feedback system (see Fig.~\ref{fig_frontend} for how the problem was implemented in the web application): 
\begin{quote}
\small
[...] "The problem reads: Two identical small balls with equal electric charge rest on a horizontal plane (see illustration). The right ball is fixed, while the left ball is movable and initially at rest far away. A piece of modeling clay with known velocity collides with the movable ball and sticks to it, so that both slide frictionlessly toward the right ball. Determine the minimum distance between the two balls." [...]
\end{quote}
The third component is a specific \textit{student solution} to the problem at hand:
\begin{quote}
\small
    [...] "A student's solution to the problem reads: \{\textit{student solution}\}." [...]
\end{quote}
The fourth component is the \textit{evidentiary scheme}, which is specific to the given problem and constitutes the largest part of the prompt. The evidentiary scheme is organized according the the knowledge types and skills students are expected to demonstrate. {\footnote{\corr{Since the focus of this study lies in providing analytical feedback on the content-level (i.e. directly observable aspects of students' physics problem solving), metacognitive knowledge — which concerns the higher-order regulation of the problem-solving process rather than its domain content — was deliberately placed outside the scope of the prompt and is left as a target for future work.}} The complete evidentiary scheme for the given problem, included in the prompt in textual form, is shown in Table~\ref{tab_evidence_scheme}. It is integrated into the overall prompt as follows:
\begin{quote}
\small
    [...] "Now provide the student with scientifically sound and physically appropriate feedback on their problem-solving approach. The feedback should address the following aspects that are relevant to a complete problem-solving approach: \{\textit{evidentiary scheme}\}." [...]
\end{quote}
The fifth and last component consists of the \textit{feedback specification}, which details pedagogical guidelines and general constraints such as feedback length. In our case, this part reads:
\begin{quote}
\small
   [...] "Check if the statements listed above are correctly included; correct them if necessary. Never reveal the full solution or final result, unless the student's solution is nearly complete and correct. Also provide the main idea or next step that would be necessary for solving the problem. Your feedback should not be longer than 100 words."
\end{quote}

\subsubsection{Costs}\label{costs}
To estimate the financial cost of a full response–feedback–response–feedback cycle, we calculated the average number of input and output tokens generated per interaction, including all backend prompting required by our system. Using the \textit{GPT-4o} API pricing as of December 2025 (USD 2.50 per million input tokens and USD 10.00 per million output tokens), the resulting average cost of a single cycle in our study was approximately USD 0.007 (0.7 cents). For perspective, at an expert human tutor rate of USD 14 per hour, this amount corresponds to only about 1.8 seconds of human tutoring time—far too little for meaningful feedback.

\begin{tablehere}
\tbl{Evidentiary scheme for the \textit{Charges} problem}
{\scriptsize
\begin{tabular}{@{}p{0.6in}p{2.6in}@{}}
\toprule
Knowledge types and skills & Evidence statements \\ \colrule

Conceptual knowledge & 
\textit{The central concepts for this physics problem are:}
\begin{itemize}[labelsep=2mm]
\item The collision between the clay and the movable ball is inelastic; hence, momentum is conserved, but kinetic energy is not.
\item As the ball (with the clay attached) approaches the fixed ball, its kinetic energy is converted into potential energy of the electric field between the balls.
\item The minimum distance is reached when all kinetic energy has been converted into potential energy.
\end{itemize} \\

Conditional knowledge & 
\textit{These concepts can be applied under the following assumptions:}
\begin{itemize}[labelsep=2mm]
\item Since the charged balls are described as small, they can be approximated as point charges, which simplifies the description of the electric field.
\item The initial potential energy of the movable ball can be neglected because it starts far away from the fixed ball.
\end{itemize} \\

Procedural knowledge & 
\textit{The problem can be solved by:}
\begin{itemize}[labelsep=2mm]
\item Dividing the problem into two parts.
\item First, using the law of conservation of momentum to calculate the joint velocity of the ball and clay after the inelastic collision.
\item Second, applying energy conservation (conversion of kinetic to potential energy) to determine the minimum distance.
\end{itemize} \\

Factual knowledge &
\textit{Relevant factual knowledge to this problem includes:}
\begin{itemize}[labelsep=2mm]
\item The momentum of a moving object is given by $p = mv$, where $m$ is its mass and $v$ its velocity.
\item The kinetic energy of a moving object is given by $E_{\text{kin}} = \tfrac{1}{2} m v^2$.
\item The potential energy of a point charge in the electric field of another (fixed) point charge is given by $E_{\text{pot}} = \tfrac{1}{4 \pi \epsilon_0} \tfrac{q_1 q_2}{r}$.
\end{itemize} \\

Mathematics skills & 
\textit{Important mathematical aspects include:}
\begin{itemize}[labelsep=2mm]
        \item Applying conservation of momentum to the collision gives $m_\text{C} v = (m_\text{C} + m_\text{B}) \tilde{v}$, where $m_\text{C}$ is the mass of the clay, $m_\text{B}$ is the mass of the ball, $v$~the velocity of the clay, and $\tilde{v}$ the velocity after the collision.
        \item Solving for $\tilde{v}$ yields $\tilde{v} = \tfrac{m_\text{C}}{m_\text{C}+m_\text{B}}v$.
        \item Energy conservation gives $E_{\text{kin}} = E_\text{pot}$, where $E_\text{kin} = \tfrac{1}{2} (m_\text{C}+m_\text{B}) \tilde{v}^2$ and $E_\text{pot} = \tfrac{q^2}{4 \pi \epsilon_0 r}$.
        \item Substituting and solving for $r$ gives the minimum distance $r = \tfrac{q^2 (m_\text{C}+m_\text{B})}{2 \pi \epsilon_0 m_\text{C}^2 v^2}$.
    \end{itemize} \ \\ 
\botrule
\end{tabular}}
\label{tab_evidence_scheme}
\end{tablehere}

\section{Evaluation}
\subsection{Study Design and Sample}
The developed feedback system\footnote{See: \url{https://wasp.leibniz-ipn.de/login}.} was tested during the first stage of the German Physics Olympiad—an annual problem-centered student competition for secondary school students across all of Germany. All students who registered for the German Physics Olympiad were invited to voluntarily use the system (and thereby participate in our study\footnote{Informed consent was assured with participants via e-mail. Participants below the age of 16 were required to register in supervision with their parents. AI-use and data protection adhere to local rules and especially the EU AI-Act and DSGVO/GDPR.}), which was advertised as a problem solving training opportunity—accessible to all participants at every time regardless of their place of residence. 

\rev{After completing each of the six problems, students could rate the usefulness and correctness of the feedback received for that problem. Since students could work on multiple problems, they could also provide multiple ratings.} \corr{Because multi-item scales \rev{based on the Technology Acceptance Model} would have substantially extended the questionnaire after each problem and risked reducing engagement or causing participant drop-out, we used a single Likert-type item.}  Regarding usefulness, after having completed a physics problem in the feedback system and having received LLM-generated feedback twice for the respective problem, students could rate their agreement with the statement: "The feedback I received for this problem helped me to better understand and work on this problem." Ratings were given on a 5-point \corr{single Likert-type item}—ranging from strongly disagree~(1) to strongly agree~(5). 
To assess students' perceived correctness of the LLM-generated feedback, students were additionally asked to rate their agreement with the statement: "The feedback appeared to be factually correct" on a 5-point \corr{single Likert-type item} as above.

\corr{\rev{A total of 38 students worked on at least one problem and rated the perceived usefulness and correctness of the feedback for the problems they attempted. As students could submit ratings for each individual problem, this resulted in 64 ratings of perceived usefulness and perceived correctness.} Each rating corresponded to one problem and was provided after the problem had been completed and feedback had been received twice. Across the different problems, we obtained 36, 8, 6, 5, 5, and 4 ratings, respectively.}
Students also had the chance to elaborate on both their ratings in the form of an open-text response. This way, we additionally received 47 written elaborations further explaining the ratings.

To determine the actual correctness of the LLM-generated feedback, the first author and a graduate student assistant independently examined the generated feedback for errors and subsequently discussed their evaluations to synthesize a consolidated judgment regarding the occurrence and nature of any errors. \corr{Feedback was rated as correct if no physical or mathematical errors occurred, whereas incorrect feedback contained at least one error. During this process, different kinds of errors were identified: calculation errors, missing or incorrect terms, incorrect physical concepts or assumptions, inappropriate solution strategies, and the misclassification of correct alternative approaches as incorrect. The open-text responses were examined for aspects that occurred repeatedly or were considered exemplary. With a larger number of open-text responses, future research could classify the responses inductively \cite{schreier_qualitative_2012} to identify categories describing how the feedback was perceived by the students.} \rev{In this study no formal qualitative analysis has been applied.}

\subsection{Results}

\subsubsection{Perceived Usefulness (RQ1)}

\revIII{The collected ratings indicate that the LLM-generated feedback was generally perceived as useful ($M = 3.6$, $SD = 1.3$ by rating; see Fig.~\ref{fig_feedback}). }
\footnote{\revIII{For perceived usefulness, the problem-specific ratings were as follows: Problem 1 (\mbox{$n = 36, M = 3.6, SD = 1.3$}), Problem 2 (\mbox{$n = 8, M = 3.4, SD = 1.6$}), Problem 3 (\mbox{$n = 6, M = 3.0, SD = 1.1$}), Problem 4 (\mbox{$n = 5, M = 4.2, SD = 0.8$}), Problem 5 (\mbox{$n = 5, M = 3.8, SD = 0.8$}), and Problem 6 (\mbox{$n = 4, M = 4.8, SD = 0.5$}).  
}}
\revIII{Because the problems were rated with unequal frequency, the rating-level mean gives greater weight to more frequently rated problems. We therefore examined the estimate's sensitivity to the aggregation method. Averaging ratings within each problem first and then across the six problems yielded essentially the same value ($M = 3.8$; the six problem means ranged from $M = 3.0$ to $M = 4.8$). Likewise, comparing the most frequently rated problem (Problem 1: \mbox{$M = 3.6$, $SD = 1.3$}) with the remaining five problems (\mbox{$M = 3.7$, $SD = 1.2$}) led to the same conclusion. The estimate was therefore insensitive to the choice of aggregation method. Because several students rated more than one problem, this estimate should be interpreted at the level of ratings rather than individual students.}}

Within students’ elaborations on their ratings, we identified ten reports stating that the feedback was very helpful for understanding and solving the problem, while in four cases the adaptivity of the feedback was described positively. \revIII{Specifically, a rating of 5/5 was further explained,} 
\begin{quote}
\small
    "I am impressed by how well the AI understood the formulas I created and the reasoning behind them, even without me defining the variables I used beforehand."\footnote{This and all following quotes were translated from German to English by the authors.}
\end{quote}
In contrast, feedback was criticized twelve times for not being sufficiently adaptive with regard to individual solutions. Specifically, a 3/5 rating came with the comment,
\begin{quote}
\small
    "My solution would have worked on the first try, but a more complicated approach was suggested to me", 
\end{quote}
and a 2/5 rating with, 
\begin{quote}
\small
    "[…] although my idea was correct, the AI said that it wasn’t quite right because I had positioned the axis differently."
\end{quote}

\subsubsection{Perceived vs. Actual Correctness (RQ2)}

\revIII{The collected ratings indicate that the LLM-generated feedback was generally perceived as highly correct  (\mbox{$M=4.4$}, \mbox{$SD=1.0$} by rating, see Fig.~\ref{fig_feedback}).}  
\footnote{\revIII{For perceived correctness, the problem-specific ratings were as follows: 
Problem 1 (\mbox{$n = 36, M = 4.2, SD = 1.0$}), 
Problem 2 (\mbox{$n = 8, M = 4.0, SD = 1.4$}), 
Problem 3 (\mbox{$n = 6, M = 4.8, SD = 0.4$}), 
Problem 4 (\mbox{$n = 5, M = 4.8, SD = 0.4$}), 
Problem 5 (\mbox{$n = 5, M = 4.6, SD = 0.9$}), and 
Problem 6 (\mbox{$n = 4, M = 5.0, SD = 0.0$}). }} \revIII{For the same reason described above, we pooled the ratings into a single estimate, which proved stable across weighting schemes (\mbox{$M = 4.4$, $SD=1.0$} by rating; \mbox{$M=4.6$} by problem, the six problem means ranged from \mbox{$M = 4.0$} to \mbox{$M = 5.0$}) and consistent with the most frequently rated problem (Problem 1: $M = 4.2$, $SD=1.0$; remaining five: $M = 4.6$, $SD=0.9$).}

\revIII{For example, the rationale for a 5/5 rating was explained in greater detail:}
\begin{quote}
\small
    "The feedback was helpful because it drew attention to the missing step that I had no longer considered necessary."
\end{quote}
\corr{Overall, an analysis of correctness by two human raters revealed} that the feedback was physically correct in 51 of 64 cases ($\approx 80\%$).
However, the remaining 13 cases ($\approx 20\%$) contained 14 minor to substantial errors. Minor errors included calculation errors, while major errors included missing or incorrect terms, incorrect physics concepts or assumptions, inappropriate solution strategies, and misclassifications of correct alternative approaches as incorrect.
\revIII{Specifically, a rating of 1/5 was further explained: }
\begin{quote}
\small
    "My solution is correct. The coordinate system is implicitly assumed such that $y$ is parallel to the electric field and $x$ is vertical to it."
\end{quote}
This is one of the two cases in which a student’s correct approach was classified as incorrect by the feedback system because the axes were defined in an unconventional way by the students.
Overall, errors in the feedback were noted by just two students in their written elaborations. \revIII{Moreover, in two cases, the feedback was described as superficial.}
 
\rev{To quantitatively assess differences in perceived correctness between cases in which the received feedback was correct and those in which it was incorrect, a two-sided Mann–Whitney $U$ test was conducted. The Mann–Whitney $U$ test is a nonparametric alternative to the independent-samples $t$-test that assesses whether two independent groups differ in their distributions without assuming normality, making it suitable for ordinal data such as the \corr{5-point single-item Likert-type scale} used to assess perceived correctness \cite{mcknightMannWhitneyTest2010}. Because students provided multiple ratings, the analysis was conducted using the rating associated with the most edited problem to ensure independence of observations. For this problem, 36 ratings were available. In 27 cases, the feedback was correct, whereas in 9 cases, the feedback was incorrect. Perceived correctness ratings were similar for students who received correct feedback (\mbox{$N_1=27$}, \mbox{$M_1=4.3$}, \mbox{$SD_1=1.0$}) and those who received incorrect feedback (\mbox{$N_2=9$}, \mbox{$M_2=4.0$}, \mbox{$SD_2=0.9$}). The test revealed no significant difference in perceived correctness ratings between students receiving correct feedback and those receiving incorrect feedback (\mbox{$U=148.5$}, \mbox{$p=0.2805$}). Although the Mann–Whitney \textit{U} test can be applied to small samples, its statistical power is limited when group sizes are small. Therefore, the non-significant result should be interpreted with caution, as it may reflect insufficient statistical power rather than the absence of a true effect.}

\vspace*{0.7cm}
\begin{figurehere}
\centering
\includegraphics[width=2.7 in]{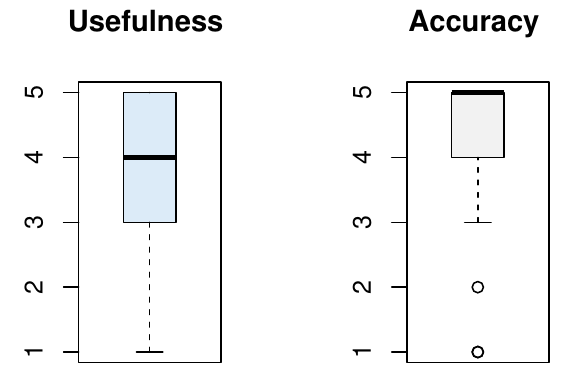}
\caption{Perceived usefulness ($M=3.6, SD=1.3$) and correctness ($M=4.4, SD=1.0$) of the LLM-generated feedback rated on 5-point single Likert-type item from $N=64$ student ratings.}
\label{fig_feedback}
\end{figurehere}


\section{Discussion \corr{and Future Directions}}

This study presented the design and implementation of an LLM-based feedback system for physics problem solving grounded in evidence-centered design and reported a first evaluation within the German Physics Olympiad. \revIII{The evaluation examined problem-related ratings of the perceived usefulness of the generated feedback (RQ1) as well as the perceived and actual correctness of the feedback (RQ2). With respect to RQ1, the LLM-generated feedback was generally perceived as useful. In cases where the feedback was not considered useful, it was often described as insufficiently adapted to individual approaches. With respect to RQ2, the feedback was, on average, perceived as highly correct.} 

However, a detailed analysis of its correctness showed that the feedback contained errors in about 20\% of cases, similar to Gupta et al.~\cite{Gupta2025BeyondFinalAnswers}. This occurred despite grounding the LLM output in ECD, which we had expected to reduce the amount of erroneous feedback. The erroneous feedback also went almost undetected by students. One may hypothesize that even high-performing Physics Olympiad participants tended to 
accept the feedback without critical reflection \cite{krupp_unreflected_2024}.
One likely reason is that LLMs present their output in the polished, expert-like language of domain specialists, thereby masking underlying errors and making them harder to detect \cite{Steyvers2025WhatLLMsKnow}. 
Thus, there exists a risk of students learning factually incorrect information.
LLM-based feedback systems should therefore make users explicitly aware that the generated feedback may contain errors and that it should not be accepted uncritically. To support this, the system should provide simple mechanisms for students to flag potentially erroneous feedback, enabling continuous monitoring and improvement. \corr{Critical oversight by educators and initiation of a design cycle is similarly important to improve upon existing weaknesses, and raise suitability of the system for a given context (here: Physics Olympiad).}

Several limitations of the system and its evaluation should be noted. 

The system relies on prompt engineering with OpenAI's \textit{GPT-4o} model following the ECD-derived problem-specific evidentiary schemes. Alternative prompting strategies, other models—including open-source LLMs—, model fine-tuning, or complementary methods such as retrieval-augmented generation were not explored and may yield more useful and accurate feedback \corr{\cite{tufino_notebooklm_2025}.}

A further limitation of the system concerns students’ criticism of the limited adaptiveness of the feedback, particularly when their solutions followed alternative—yet valid—approaches not represented in the problem-specific evidentiary schemes (see e.g., Table~\ref{tab_evidence_scheme}). The current ECD-based approach is not well suited to handle such variability and implicitly pushes students toward a single canonical solution path, potentially flagging viable alternatives as incorrect. One way forward is to iteratively integrate additional solution paths into the evidentiary schemes; alternatively, an anomaly-detection layer could be implemented. In this approach, the system would first determine whether a student’s solution matches the canonical solution path: if so, ECD-grounded feedback is generated; if not, the system falls back on general LLM reasoning, with the caveat that such feedback is more prone to errors because it is not anchored in an underlying model solution.

A further development perspective to improve adaptivity of LLM-based feedback systems is the integration of a so-called student model \cite{chrysafiadiStudentModelingApproaches2013} which should contain information about students' mastery of the introduced knowledge types and skills (see Table \ref{tab:evidence_statements}). In our system, the arrangement of problems was static and did not follow a curricular model. This constitutes a limitation with respect to enabling continuous and targeted deliberate practice, which is essential for the development of problem-solving abilities \cite{Ericsson.1998}. Combining the ECD approach (inner loop) with an outer loop would make it possible to adapt both the selection of subsequent problems and the mode of feedback to students' current mastery of problem-solving abilities. In this way, the adaptivity of the overall feedback system might be improved.

Regarding the system evaluation, one limitation is that it assessed the perceived usefulness, perceived correctness, and actual correctness of the feedback, but not its actual impact on learning. Future research should investigate whether such feedback leads to measurable learning gains and improved problem-solving performance. \corr{Another limitation is that, in order to maintain usability for the students, perceived usefulness and perceived correctness were each measured using a single Likert-type item rather than a validated scale. \rev{Moreover, mean values across all problems were used because most problems had not been rated by students frequently enough to allow for comparisons of mean values across the different problems.}}



\corr{Despite these limitations, improving the system appears worthwhile. It provides immediate feedback on complex physics problem-solving tasks with minimal delay and at very low cost, \revIII{and the LLM-generated feedback was generally rated as useful.} As outlined in Subsection~\ref{costs}, generating feedback twice for a single problem-solving approach incurred costs of only USD~0.007 (0.7 cents). Thus, although the system requires further improvement in terms of adaptivity, reliability, and demonstrated learning impact, it offers a scalable basis for providing timely feedback on complex problem-solving processes.}

\section{Conclusions}

By using ECD as the foundation for generating LLM-based feedback, we were able to produce analytical feedback for students' physics problem-solving approaches. The ECD approach is particularly practical, as new problems only require specification of the underlying knowledge types, a process already indirectly carried out during problem development. 

More broadly, the ECD approach demonstrates how LLM-based feedback for physics problem solving can be systematically grounded in domain knowledge, rather than relying on unguided LLM reasoning alone. However, as our findings show, this grounding alone does not guarantee accurate feedback. \revIII{The feedback was perceived as useful and highly correct.}; yet, despite its ECD grounding, 20\% of the feedback contained minor to substantial errors, which were rarely detected by the students. This finding points to a broader challenge for AI-assisted physics education: physics problem solving demands precise reasoning, and if students uncritically adopt flawed feedback, the very abilities the feedback aims to develop may be undermined. As it does not seem possible to completely avoid errors, the question arises as to how students can be prepared to critically evaluate AI-generated content—a competence that is becoming increasingly relevant for physics learning beyond this specific feedback system. 

\corr{\revIII{In addition, the ratings indicated the limited adaptivity of the system.} Thus, while the ECD-based approach provides a strong starting point for generating LLM-based feedback for complex activities such as physics problem solving, further research is needed to reduce the rate of erroneous feedback and improve adaptivity. Retrieval-augmented generation and the flagging of potentially erroneous feedback may improve correctness, while integrating alternative solution paths and incorporating a student model may enhance the adaptivity of the generated feedback.}

\corr{We echo the call by Dong et al.\cite{dongExaminingEffectArtificial2025} for further research related to AI in education, however, we emphasize that educational processes for a large part are domain-specific. On the one hand, if LLM-based automated feedback systems should be developed in specific context such as the Physics Olympiad, domain experts need to oversee the human-AI-interactions. On the other hand, if more encompassing learning systems with AI integration should be developed, multi-disciplinary expertise will be required to genuinely evaluate the learning effectiveness of such systems. These systems are too consequential for students' learning processes to let them be developed by single programmers who might be able to setup an LLM that mimics authentic dialogues, but which fails to scaffold conceptual change, help represent a problem in meaningful ways, or provide guidance on approaching physics problems in general.}

\section*{Funding}
This work was supported by the Klaus-Tschira-Stiftung (project \textit{WasP}) under Grant No. 00.001.2023.

\section*{Acknowledgments}
ChatGPT (OpenAI) was used for language editing to improve clarity and style. The authors reviewed and take full responsibility for the final content of the manuscript.

\section*{ORCID}

\noindent Holger Maus - \par \url{https://orcid.org/0009-0008-8555-1405}

\section*{References}
\bibliographystyle{ws-tpe}
\bibliography{ref}

@article{BingRedish2009,
  author  = {Bing, Thomas J. and Redish, Edward F.},
  title   = {Analyzing problem solving using math in physics: Epistemological framing via warrants},
  journal = {Physical Review Special Topics - Physics Education Research},
  volume  = {5},
  pages   = {020108},
  year    = {2009}
}

@article{Ericsson.1998,
 author = {Ericsson, K. Anders},
 year = {1998},
 title = {Scientific study of expert levels of performance: General implications for optimal learning and creativity},
 pages = {75--110},
  volume = {9},
 number = {9},
 journal = {High Ability Studies}
}

@article{Kim.2002,
 author = {Kim, Eunsook and Pak, Sung-Jae},
 year = {2002},
 title = {Students do not overcome conceptual difficulties after solving 1000 traditional problems},
 pages = {759--765},
 volume = {70},
 number = {7},
 journal = {American Journal of Physics},
 doi = {10.1119/1.1484151}
}

@article{Tegmark.2008,
 author = {Tegmark, Max},
 year = {2008},
 title = {The Mathematical Universe},
 pages = {101--150},
 volume = {38},
 number = {2},
 issn = {0015-9018},
 journal = {Foundations of Physics},
 doi = {10.1007/s10701-007-9186-9}
}

@article{Witte1972PhaseTheorem,
  author  = {Witte, Eberhard},
  title   = {Field Research on Complex Decision-Making Processes—The Phase Theorem},
  journal = {International Studies of Management \& Organization},
  year    = {1972},
  volume  = {2},
  number  = {2},
  pages   = {156--182},
  doi     = {10.1080/00208825.1972.11656117}
}

@article{tschisgaleExploringSequentialStructure2025,
  title = {Exploring the Sequential Structure of Students' Physics Problem-Solving Approaches Using Process Mining and Sequence Analysis},
  author = {Tschisgale, Paul and Kubsch, Marcus and Wulff, Peter and Petersen, Stefan and Neumann, Knut},
  year = 2025,
  month = jan,
  journal = {Physical Review Physics Education Research},
  volume = {21},
  number = {1},
  pages = {010111},
  issn = {2469-9896},
  doi = {10.1103/PhysRevPhysEducRes.21.010111},
  urldate = {2025-02-03},
  langid = {english}
}

@article{Shuster2021_RAG_Hallucination,
  title        = {Retrieval Augmentation Reduces Hallucination in Conversation},
  author       = {Shuster, Kurt and Poff, Samuel and Chen, Michel and Kiela, Douwe and Weston, Jason},
  journal      = {arXiv preprint},
  year         = {2021},
  note         = {{arXiv:2104.07567}},
  url          = {https://arxiv.org/abs/2104.07567}
}

@article{chrysafiadiStudentModelingApproaches2013,
  title = {Student Modeling Approaches: {{A}} Literature Review for the Last Decade},
  shorttitle = {Student Modeling Approaches},
  author = {Chrysafiadi, Konstantina and Virvou, Maria},
  year = 2013,
  month = sep,
  journal = {Expert Systems with Applications},
  volume = {40},
  number = {11},
  pages = {4715--4729},
  issn = {09574174},
  doi = {10.1016/j.eswa.2013.02.007},
  urldate = {2022-03-22},
  langid = {english}
}

@incollection{mcknightMannWhitneyTest2010,
  title = {Mann-{{Whitney U}} Test},
  shorttitle = {Mann-{{Whitney}}},
  booktitle = {The {{Corsini Encyclopedia}} of {{Psychology}}},
  author = {McKnight, Patrick E. and Najab, Julius},
  editor = {Weiner, Irving B. and Craighead, W. Edward},
  year = 2010,
  month = jan,
  edition = {1},
  pages = {1--1},
  publisher = {Wiley},
  doi = {10.1002/9780470479216.corpsy0524},
  urldate = {2025-03-10},
  copyright = {http://doi.wiley.com/10.1002/tdm\_license\_1.1},
  isbn = {978-0-470-17024-3 978-0-470-47921-6},
  langid = {english}
}

@article{federiakinPromptEngineeringNew2024,
  title = {Prompt Engineering as a New 21st Century Skill},
  author = {Federiakin, Denis and Molerov, Dimitri and {Zlatkin-Troitschanskaia}, Olga and Maur, Andreas},
  year = 2024,
  month = nov,
  journal = {Frontiers in Education},
  volume = {9},
  publisher = {Frontiers Media SA},
  issn = {2504-284X},
  doi = {10.3389/feduc.2024.1366434},
  urldate = {2025-07-16},
  copyright = {https://creativecommons.org/licenses/by/4.0/}
}

@article{tschisgale2024science,
  title={Are science competitions meeting their intentions? a case study on affective and cognitive predictors of success in the Physics Olympiad},
  author={Tschisgale, Paul Leon and Steegh, Anneke and Petersen, Stefan and Kubsch, Marcus and Wulff, Peter and Neumann, Knut},
  journal={Disciplinary and Interdisciplinary Science Education Research},
  volume={6},
  number={1},
  pages={10},
  year={2024},
  publisher={Springer}
}

@article{PhysRevPhysEducRes.21.010111,
  title = {Exploring the sequential structure of students' physics problem-solving approaches using process mining and sequence analysis},
  author = {Tschisgale, Paul and Kubsch, Marcus and Wulff, Peter and Petersen, Stefan and Neumann, Knut},
  journal = {Phys. Rev. Phys. Educ. Res.},
  volume = {21},
  issue = {1},
  pages = {010111},
  numpages = {21},
  year = {2025},
  month = {Jan},
  publisher = {American Physical Society},
  doi = {10.1103/PhysRevPhysEducRes.21.010111},
  url = {https://link.aps.org/doi/10.1103/PhysRevPhysEducRes.21.010111}
}

@article{kortemeyer_could_2023,
	title = {Could an {Artificial}-{Intelligence} agent pass an introductory physics course?},
	volume = {19},
	issn = {2469-9896},
	doi = {10.1103/PhysRevPhysEducRes.19.010132},
	number = {1},
	urldate = {2025-07-07},
	journal = {Physical Review Physics Education Research},
	author = {Kortemeyer, Gerd},
	month = may,
	year = {2023},
}

@article{Rafner.2021,
 author = {Rafner, Janet and Dellermann, Dominik and Hjorth, Arthur and Veraszt{\'o}, D{\'o}ra and Kampf, Constance and Mackay, Wendy and Sherson, Jacob},
 year = {2021},
 title = {Deskilling, Upskilling, and Reskilling: a Case for Hybrid Intelligence},
 pages = {24--39},
 volume = {1},
 number = {2},
 issn = {2747-5174},
 journal = {Morals {\&} Machines},
 doi = {10.5771/2747-5174-2021-2-24}
}

@article{tschisgale_towards_2024,
	title = {Towards a more individualised support of science competition participants – identification and examination of participant profiles based on cognitive and affective characteristics},
	volume = {46},
	copyright = {http://creativecommons.org/licenses/by/4.0/},
	issn = {0950-0693, 1464-5289},
	doi = {10.1080/09500693.2023.2300147},
	language = {en},
	number = {16},
	urldate = {2025-07-07},
	journal = {International Journal of Science Education},
	author = {Tschisgale, Paul and Steegh, Anneke and Kubsch, Marcus and Petersen, Stefan and Neumann, Knut},
	month = nov,
	year = {2024},
	pages = {1757--1781},
}

@article{burkholder_characterizing_2020,
	title = {Characterizing the mathematical problem-solving strategies of transitioning novice physics students},
	volume = {16},
	issn = {2469-9896},
	doi = {10.1103/PhysRevPhysEducRes.16.020134},
	language = {en},
	number = {2},
	urldate = {2025-07-29},
	journal = {Physical Review Physics Education Research},
	author = {Burkholder, Eric and Blackmon, Lena and Wieman, Carl},
	month = nov,
	year = {2020},
	pages = {020134},
}

@article{friege_types_2006,
	title = {Types and {Qualities} of {Knowledge} and their {Relations} to {Problem} {Solving} in {Physics}},
	volume = {4},
	copyright = {http://www.springer.com/tdm},
	issn = {1571-0068, 1573-1774},
	doi = {10.1007/s10763-005-9013-8},
	language = {en},
	number = {3},
	urldate = {2025-07-29},
	journal = {International Journal of Science and Mathematics Education},
	author = {Friege, Gunnar and Lind, Gunter},
	month = nov,
	year = {2006},
	pages = {437--465},
}

@inproceedings{helal_when_2024,
	address = {Cairo, Egypt},
	title = {When the {Robotic} {Maths} {Tutor} is {Wrong} - {Can} {Children} {Identify} {Mistakes} {Generated} by {ChatGPT}?},
	copyright = {https://doi.org/10.15223/policy-029},
	isbn = {979-8-3503-8507-6},
	doi = {10.1109/AIRC61399.2024.10672220},
	urldate = {2025-08-05},
	booktitle = {2024 5th {International} {Conference} on {Artificial} {Intelligence}, {Robotics} and {Control} ({AIRC})},
	publisher = {IEEE},
	author = {Helal, Manal and Holthaus, Patrick and Wood, Luke and Velmurugan, Vignesh and Lakatos, Gabriella and Moros, Silvia and Amirabdollahian, Farshid},
	month = apr,
	year = {2024},
	pages = {83--90},
}

@article{avila_using_2024,
	title = {Using {ChatGPT} for {Teaching} {Physics}},
	volume = {62},
	issn = {0031-921X, 1943-4928},
	doi = {10.1119/5.0227132},
	number = {6},
	urldate = {2025-05-16},
	journal = {The Physics Teacher},
	author = {Avila, Karina E. and Steinert, Steffen and Ruzika, Stefan and Kuhn, Jochen and Küchemann, Stefan},
	month = sep,
	year = {2024},
	pages = {536--537},
}

@book{bransford_how_2000,
	address = {Washington, DC},
	title = {How {People} {Learn}},
	volume = {11},
	publisher = {National Academy Press},
	author = {Bransford, John D. and Brown, Ann L. and Cocking, Rodney R.},
	year = {2000},
}

@article{chen_grading_2025,
	title = {Grading {Explanations} of {Problem}-{Solving} {Process} and {Generating} {Feedback} {Using} {Large} {Language} {Models} at {Human}-{Level} {Accuracy}},
	volume = {21},
	issn = {2469-9896},
	doi = {10.1103/PhysRevPhysEducRes.21.010126},
	number = {1},
	urldate = {2025-04-15},
	journal = {Physical Review Physics Education Research},
	author = {Chen, Zhongzhou and Wan, Tong},
	month = mar,
	year = {2025},
	pages = {010126},
}

@article{cheng_science_2025,
	title = {Science {Education} in the {Age} of {Artificial} {Intelligence}: {Opportunities}, {Challenges}, and {Research}},
	volume = {18},
	copyright = {https://ieeexplore.ieee.org/Xplorehelp/downloads/license-information/IEEE.html},
	issn = {1939-1382, 2372-0050},
	shorttitle = {Science {Education} in the {Age} of {Artificial} {Intelligence}},
	doi = {10.1109/TLT.2025.3575030},
	urldate = {2025-08-05},
	journal = {IEEE Transactions on Learning Technologies},
	author = {Cheng, May Hung May and Wan, Zhi Hong},
	year = {2025},
	pages = {635--638},
}

@article{docktor_assessing_2016,
	title = {Assessing {Student} {Written} {Problem} {Solutions}: {A} {Problem}-{Solving} {Rubric} with {Application} to {Introductory} {Physics}},
	volume = {12},
	shorttitle = {Assessing {Student} {Written} {Problem} {Solutions}},
	doi = {10.1103/PhysRevPhysEducRes.12.010130},
	number = {1},
	urldate = {2021-05-07},
	journal = {Physical Review Physics Education Research},
	author = {Docktor, Jennifer L. and Dornfeld, Jay and Frodermann, Evan and Heller, Kenneth and Hsu, Leonardo and Jackson, Koblar Alan and Mason, Andrew and Ryan, Qing X. and Yang, Jie},
	month = may,
	year = {2016},
}

@article{fan_beware_2025,
	title = {Beware of {Metacognitive} {Laziness}: {Effects} of {Generative} {Artificial} {Intelligence} on {Learning} {Motivation}, {Processes}, and {Performance}},
	volume = {56},
	issn = {0007-1013, 1467-8535},
	shorttitle = {Beware of {Metacognitive} {Laziness}},
	doi = {10.1111/bjet.13544},
	number = {2},
	urldate = {2025-04-28},
	journal = {British Journal of Educational Technology},
	author = {Fan, Yizhou and Tang, Luzhen and Le, Huixiao and Shen, Kejie and Tan, Shufang and Zhao, Yueying and Shen, Yuan and Li, Xinyu and Gašević, Dragan},
	month = mar,
	year = {2025},
	pages = {489--530},
}

@article{frey_teaching_2022,
	title = {Teaching {Discipline}-{Based} {Problem} {Solving}},
	volume = {21},
	issn = {1931-7913},
	doi = {10.1187/cbe.22-02-0030},
	number = {2},
	urldate = {2022-12-16},
	journal = {CBE—Life Sciences Education},
	author = {Frey, Regina F. and Brame, Cynthia J. and Fink, Angela and Lemons, Paula P.},
	editor = {Wolfson, Adele},
	month = jun,
	year = {2022},
}

@article{gaigher_exploring_2007,
	title = {Exploring the {Development} of {Conceptual} {Understanding} through {Structured} {Problem}-solving in {Physics}},
	volume = {29},
	issn = {0950-0693, 1464-5289},
	doi = {10.1080/09500690600930972},
	number = {9},
	urldate = {2023-11-23},
	journal = {International Journal of Science Education},
	author = {Gaigher, E. and Rogan, J. M. and Braun, M. W. H.},
	month = jul,
	year = {2007},
	pages = {1089--1110},
}

@misc{OpenAI2023GPT4TechReport,
  title         = {{GPT-4} Technical Report},
  author        = {OpenAI},
  year          = {2023},
  month         = {mar},
  eprint        = {2303.08774},
  archivePrefix = {arXiv},
  primaryClass  = {cs.CL},
  doi           = {10.48550/arXiv.2303.08774},
  url           = {https://arxiv.org/abs/2303.08774}
}

@article{kortemeyer_boiling-frog_2026,
	title = {The {Boiling}-{Frog} {Problem} of {Physics} {Education}},
	volume = {64},
	issn = {0031-921X, 1943-4928},
	url = {https://pubs.aip.org/pte/article/64/1/8/3375731/The-Boiling-Frog-Problem-of-Physics-Education},
	doi = {10.1119/5.0296601},
	language = {en},
	number = {1},
	urldate = {2026-05-22},
	journal = {The Physics Teacher},
	author = {Kortemeyer, Gerd},
	month = jan,
	year = {2026},
	pages = {8--12},
}

@article{Steyvers2025WhatLLMsKnow,
  title   = {What large language models know and what people think they know},
  author  = {Steyvers, Mark and Tejeda, Heliodoro and Kumar, Aakriti and Belem, Catarina and Karny, Sheer and Hu, Xinyue and Mayer, Lukas W. and Smyth, Padhraic},
  journal = {Nature Machine Intelligence},
  year    = {2025},
  volume  = {7},
  pages   = {221--231},
  doi     = {10.1038/s42256-024-00976-7},
}

@book{Summerfield.2025,
 author = {Summerfield, Christopher},
 year = {2025},
 title = {These strange new minds: How AI learned to talk and what it means},
 address = {London},
 publisher = {{Penguin Viking}},
 isbn = {9780241694657}
}

@article{Bastani.2025,
 author = {Bastani, Hamsa and Bastani, Osbert and Sungu, Alp and Ge, Haosen and Kabakc{\i}, {\"O}zge and Mariman, Rei},
 year = {2025},
 title = {Generative AI without guardrails can harm learning: Evidence from high school mathematics},
 pages = {e2422633122},
 volume = {122},
 number = {26},
 issn = {1091-6490},
 journal = {Proceedings of the National Academy of Sciences of the United States of America},
 doi = {10.1073/pnas.2422633122}
}

@article{cheng_sycophantic_2026,
	title = {Sycophantic {AI} decreases prosocial intentions and promotes dependence},
	volume = {391},
	issn = {0036-8075, 1095-9203},
	url = {https://www.science.org/doi/10.1126/science.aec8352},
	doi = {10.1126/science.aec8352},
	language = {en},
	number = {6792},
	urldate = {2026-05-22},
	journal = {Science},
	author = {Cheng, Myra and Lee, Cinoo and Khadpe, Pranav and Yu, Sunny and Han, Dyllan and Jurafsky, Dan},
	month = mar,
	year = {2026},
	pages = {eaec8352},
}

@article{hattie_power_2007,
	title = {The {Power} of {Feedback}},
	volume = {77},
	copyright = {https://journals.sagepub.com/page/policies/text-and-data-mining-license},
	issn = {0034-6543, 1935-1046},
	doi = {10.3102/003465430298487},
	number = {1},
	urldate = {2025-08-15},
	journal = {Review of Educational Research},
	author = {Hattie, John and Timperley, Helen},
	month = mar,
	year = {2007},
	pages = {81--112},
}

@article{jescovitch_comparison_2021,
	title = {Comparison of {Machine} {Learning} {Performance} {Using} {Analytic} and {Holistic} {Coding} {Approaches} across {Constructed} {Response} {Assessments} {Aligned} to a {Science} {Learning} {Progression}},
	volume = {30},
	issn = {1059-0145, 1573-1839},
	doi = {10.1007/s10956-020-09858-0},
	number = {2},
	urldate = {2023-01-17},
	journal = {Journal of Science Education and Technology},
	author = {Jescovitch, Lauren N. and Scott, Emily E. and Cerchiara, Jack A. and Merrill, John and Urban-Lurain, Mark and Doherty, Jennifer H. and Haudek, Kevin C.},
	month = apr,
	year = {2021},
	pages = {150--167},
}

@article{kortemeyer_multilingual_2025,
	title = {Multilingual {Performance} of a {Multimodal} {Artificial} {Intelligence} {System} on {Multisubject} {Physics} {Concept} {Inventories}},
	volume = {21},
	copyright = {https://creativecommons.org/licenses/by/4.0/},
	issn = {2469-9896},
	doi = {10.1103/98hg-rkrf},
	number = {2},
	urldate = {2025-07-09},
	journal = {Physical Review Physics Education Research},
	author = {Kortemeyer, Gerd and Babayeva, Marina and Polverini, Giulia and Widenhorn, Ralf and Gregorcic, Bor},
	month = jul,
	year = {2025},
}

@misc{kosmyna_your_2025,
	title = {Your {Brain} on {ChatGPT}: {Accumulation} of {Cognitive} {Debt} {When} {Using} an {AI} {Assistant} for {Essay} {Writing} {Task}},
	copyright = {Creative Commons Attribution Non Commercial Share Alike 4.0 International},
	shorttitle = {Your {Brain} on {ChatGPT}},
	urldate = {2025-07-02},
	publisher = {arXiv},
	author = {Kosmyna, Nataliya and Hauptmann, Eugene and Yuan, Ye Tong and Situ, Jessica and Liao, Xian-Hao and Beresnitzky, Ashly Vivian and Braunstein, Iris and Maes, Pattie},
	year = {2025},
	doi = {10.48550/ARXIV.2506.08872},
}

@incollection{krupp_unreflected_2024,
	title = {Unreflected {Acceptance} – {Investigating} the {Negative} {Consequences} of {ChatGPT}-assisted {Problem} {Solving} in {Physics} {Education}},
	copyright = {https://creativecommons.org/licenses/by-nc/4.0/},
	isbn = {978-1-64368-522-9},
	urldate = {2025-04-16},
	booktitle = {Frontiers in {Artificial} {Intelligence} and {Applications}},
	publisher = {IOS Press},
	author = {Krupp, Lars and Steinert, Steffen and Kiefer-Emmanouilidis, Maximilian and Avila, Karina E. and Lukowicz, Paul and Kuhn, Jochen and Küchemann, Stefan and Karolus, Jakob},
	editor = {Lorig, Fabian and Tucker, Jason and Dahlgren Lindström, Adam and Dignum, Frank and Murukannaiah, Pradeep and Theodorou, Andreas and Yolum, Pınar},
	month = jun,
	year = {2024},
	doi = {10.3233/FAIA240195},
}

@article{kubsch_toward_2022,
	title = {Toward {Learning} {Progression} {Analytics} — {Developing} {Learning} {Environments} for the {Automated} {Analysis} of {Learning} {Using} {Evidence} {Centered} {Design}},
	volume = {7},
	issn = {2504-284X},
	doi = {10.3389/feduc.2022.981910},
	urldate = {2024-04-04},
	journal = {Frontiers in Education},
	author = {Kubsch, Marcus and Czinczel, Berrit and Lossjew, Jannik and Wyrwich, Tobias and Bednorz, David and Bernholt, Sascha and Fiedler, Daniela and Strauß, Sebastian and Cress, Ulrike and Drachsler, Hendrik and Neumann, Knut and Rummel, Nikol},
	month = aug,
	year = {2022},
	pages = {981910},
}

@article{leonard_using_1996,
	title = {Using {Qualitative} {Problem}-solving {Strategies} to {Highlight} the {Role} of {Conceptual} {Knowledge} in {Solving} {Problems}},
	volume = {64},
	issn = {0002-9505},
	doi = {10.1119/1.18409},
	number = {12},
	urldate = {2021-05-07},
	journal = {American Journal of Physics},
	author = {Leonard, William J. and Dufresne, Robert J. and Mestre, Jose P.},
	month = dec,
	year = {1996},
	note = {Publisher: American Association of Physics Teachers},
	pages = {1495--1503},
}

@article{mislevy_brief_2003,
	title = {A {Brief} {Introduction} to {Evidence}-{Centered} {Design}},
	volume = {2003},
	copyright = {http://onlinelibrary.wiley.com/termsAndConditions\#vor},
	issn = {2330-8516, 2330-8516},
	doi = {10.1002/j.2333-8504.2003.tb01908.x},
	number = {1},
	urldate = {2025-08-21},
	journal = {ETS Research Report Series},
	author = {Mislevy, Robert J. and Almond, Russell G. and Lukas, Janice F.},
	month = jun,
	year = {2003},
}

@article{poldner_assessing_2014,
	title = {Assessing {Student} {Teachers}' {Reflective} {Writing} through {Quantitative} {Content} {Analysis}},
	volume = {37},
	issn = {0261-9768, 1469-5928},
	doi = {10.1080/02619768.2014.892479},
	number = {3},
	urldate = {2025-08-15},
	journal = {European Journal of Teacher Education},
	author = {Poldner, Eric and Van Der Schaaf, Marieke and Simons, P. Robert-Jan and Van Tartwijk, Jan and Wijngaards, Guus},
	month = jul,
	year = {2014},
	pages = {348--373},
}

@inproceedings{smith_toward_1988,
	address = {New Orleans, LA},
	title = {Toward a {Unified} {Theory} of {Problem} {Solving}: {A} {View} from {Biology}},
	shorttitle = {Toward a {Unified} {Theory} of {Problem} {Solving}},
	urldate = {2021-02-15},
	booktitle = {Annual {Meeting} of the {American} {Educational} {Research} {Association}},
	author = {Smith, Mike U.},
	month = apr,
	year = {1988},
}

@article{tschisgale_evaluating_2025,
	title = {Evaluating {GPT}- and {Reasoning}-{Based} {Large} {Language} {Models} on {Physics} {Olympiad} {Problems}: {Surpassing} {Human} {Performance} and {Implications} for {Educational} {Assessment}},
	volume = {21},
	issn = {2469-9896},
	shorttitle = {Evaluating {GPT}- and {Reasoning}-{Based} {Large} {Language} {Models} on {Physics} {Olympiad} {Problems}},
	doi = {10.1103/6fmx-bsnl},
	number = {2},
	urldate = {2025-08-14},
	journal = {Physical Review Physics Education Research},
	author = {Tschisgale, Paul and Maus, Holger and Kieser, Fabian and Kroehs, Ben and Petersen, Stefan and Wulff, Peter},
	month = aug,
	year = {2025},
	pages = {020115},
}

@article{xavier_empowering_2025,
	title = {Empowering {Instructors} with {AI}: {Evaluating} the {Impact} of an {AI}-driven {Feedback} {Tool} in {Learning} {Analytics}},
	volume = {18},
	copyright = {https://creativecommons.org/licenses/by/4.0/legalcode},
	issn = {1939-1382, 2372-0050},
	shorttitle = {Empowering {Instructors} {With} {AI}},
	doi = {10.1109/TLT.2025.3562379},
	urldate = {2025-08-05},
	journal = {IEEE Transactions on Learning Technologies},
	author = {Xavier, Cleon and Rodrigues, Luiz and Costa, Newarney and Neto, Rodrigues and Alves, Gabriel and Falcão, Taciana Pontual and Gašević, Dragan and Mello, Rafael Ferreira},
	year = {2025},
	pages = {498--512},
}

@article{yin_using_2024,
	title = {Using a {Chatbot} to {Provide} {Formative} {Feedback}: {A} {Longitudinal} {Study} of {Intrinsic} {Motivation}, {Cognitive} {Load}, and {Learning} {Performance}},
	volume = {17},
	copyright = {https://ieeexplore.ieee.org/Xplorehelp/downloads/license-information/IEEE.html},
	issn = {1939-1382, 2372-0050},
	shorttitle = {Using a {Chatbot} to {Provide} {Formative} {Feedback}},
	doi = {10.1109/TLT.2024.3364015},
	urldate = {2025-08-21},
	journal = {IEEE Transactions on Learning Technologies},
	author = {Yin, Jiaqi and Goh, Tiong-Thye and Hu, Yi},
	year = {2024},
	pages = {1378--1389},
}

@misc{yuHiPhOHowFar2025,
  title = {{{HiPhO}}: {{How}} Far Are ({{M}}){{LLMs}} from Humans in the Latest High School Physics {{Olympiad}} Benchmark?},
  shorttitle = {{{HiPhO}}},
  author = {Yu, Fangchen and Wan, Haiyuan and Cheng, Qianjia and Zhang, Yuchen and Chen, Jiacheng and Han, Fujun and Wu, Yulun and Yao, Junchi and Hu, Ruilizhen and Ding, Ning and Cheng, Yu and Chen, Tao and Bai, Lei and Zhou, Dongzhan and Luo, Yun and Cui, Ganqu and Ye, Peng},
  year = {2025},
  publisher = {arXiv},
  doi = {10.48550/ARXIV.2509.07894},
  urldate = {2025-09-24},
  copyright = {Creative Commons Attribution 4.0 International}
}

@misc{mokUsingAILarge2024,
  title = {Using {{AI}} Large Language Models for Grading in Education: {{A}} Hands-on Test for Physics},
  shorttitle = {Using {{AI Large Language Models}} for {{Grading}} in {{Education}}},
  author = {Mok, Ryan and Akhtar, Faraaz and Clare, Louis and Li, Christine and Ida, Jun and Ross, Lewis and Campanelli, Mario},
  year = {2024},
  publisher = {arXiv},
  doi = {10.48550/ARXIV.2411.13685},
  urldate = {2025-04-11},
  copyright = {Creative Commons Attribution 4.0 International}
}

@article{kortemeyerGradingAssistanceHandwritten2024a,
  title = {Grading Assistance for a Handwritten Thermodynamics Exam Using Artificial Intelligence: {{An}} Exploratory Study},
  shorttitle = {Grading Assistance for a Handwritten Thermodynamics Exam Using Artificial Intelligence},
  author = {Kortemeyer, Gerd and N{\"o}hl, Julian and Onishchuk, Daria},
  year = {2024},
  month = nov,
  journal = {Physical Review Physics Education Research},
  volume = {20},
  number = {2},
  pages = {020144},
  issn = {2469-9896},
  doi = {10.1103/PhysRevPhysEducRes.20.020144},
  urldate = {2025-04-16},
  langid = {english}
}

@incollection{kieserUsingLargeLanguage2024,
  title = {Using Large Language Models to Probe Cognitive Constructs, Augment Data, and Design Instructional Materials},
  booktitle = {Machine {{Learning}} in {{Educational Sciences}}},
  author = {Kieser, Fabian and Wulff, Peter},
  editor = {Khine, Myint Swe},
  year = {2024},
  pages = {293--313},
  publisher = {Springer Nature Singapore},
  address = {Singapore},
  urldate = {2024-03-12},
  isbn = {978-981-99-9378-9 978-981-99-9379-6},
  langid = {english}
}

@misc{Gupta2025BeyondFinalAnswers,
  title         = {Beyond Final Answers: Evaluating Large Language Models for Math Tutoring},
  author        = {Gupta, Adit and Reddig, Jennifer and Cal\`o, Tommaso and Weitekamp, Daniel and MacLellan, Christopher J.},
  year          = {2025},
  month         = {feb},
  eprint        = {2503.16460},
  archivePrefix = {arXiv},
  primaryClass  = {cs.HC},
  doi           = {10.48550/arXiv.2503.16460},
}

@misc{sirnoorkarFeedbackThatClicks2025,
  title = {Feedback That Clicks: {{Introductory}} Physics Students' Valued Features in {{AI}} Feedback Generated from Self-Crafted and Engineered Prompts},
  shorttitle = {Feedback {{That Clicks}}},
  author = {Sirnoorkar, Amogh and Rebello, N. Sanjay},
  year = {2025},
  publisher = {arXiv},
  doi = {10.48550/ARXIV.2509.08516},
  urldate = {2025-09-23},
  copyright = {Creative Commons Attribution 4.0 International}
}

@article{wanExploringGenerativeAI2024,
  title = {Exploring Generative {{AI}} Assisted Feedback Writing for Students' Written Responses to a Physics Conceptual Question with Prompt Engineering and Few-Shot Learning},
  author = {Wan, Tong and Chen, Zhongzhou},
  year = {2024},
  month = jun,
  journal = {Physical Review Physics Education Research},
  volume = {20},
  number = {1},
  pages = {010152},
  issn = {2469-9896},
  doi = {10.1103/PhysRevPhysEducRes.20.010152},
  urldate = {2024-06-14},
  langid = {english}
}

@article{dongExaminingEffectArtificial2025,
  title = {Examining the Effect of Artificial Intelligence in Relation to Students' Academic Achievement: {{A}} Meta-Analysis},
  shorttitle = {Examining the Effect of Artificial Intelligence in Relation to Students' Academic Achievement},
  author = {Dong, Liu and Tang, Xiuxiu and Wang, Xiyu},
  year = {2025},
  month = jun,
  journal = {Computers and Education: Artificial Intelligence},
  volume = {8},
  pages = {100400},
  issn = {2666920X},
  doi = {10.1016/j.caeai.2025.100400},
  urldate = {2025-09-24},
  langid = {english}
}

@article{tufino_notebooklm_2025,
	title = {{NotebookLM} as a {Socratic} {Physics} {Tutor}: {Design} and {Preliminary} {Observations} of a {RAG}-{Based} {Tool}},
	volume = {07},
	issn = {2661-3395, 2661-3409},
	shorttitle = {{NotebookLM} as a {Socratic} {Physics} {Tutor}},
	url = {https://www.worldscientific.com/doi/10.1142/S2661339525500143},
	doi = {10.1142/S2661339525500143},
	language = {en},
	urldate = {2026-05-22},
	journal = {The Physics Educator},
	author = {Tufino, Eugenio},
	month = jan,
	year = {2025},
	pages = {2550014},
}

@article{kestinAITutoringOutperforms2025,
  title = {{{AI}} Tutoring Outperforms In-Class Active Learning: An {{RCT}} Introducing a Novel Research-Based Design in an Authentic Educational Setting},
  shorttitle = {{{AI}} Tutoring Outperforms In-Class Active Learning},
  author = {Kestin, Greg and Miller, Kelly and Klales, Anna and Milbourne, Timothy and Ponti, Gregorio},
  year = {2025},
  month = jun,
  journal = {Scientific Reports},
  volume = {15},
  number = {1},
  pages = {17458},
  issn = {2045-2322},
  doi = {10.1038/s41598-025-97652-6},
  urldate = {2025-09-24},
  langid = {english}
}

@book{schreier_qualitative_2012,
	address = {1 Oliver's Yard, 55 City Road  London  EC1Y 1SP},
	title = {Qualitative {Content} {Analysis} in {Practice}},
	isbn = {978-1-84920-593-1 978-1-5296-8257-1},
	url = {https://methods.sagepub.com/book/qualitative-content-analysis-in-practice},
	doi = {10.4135/9781529682571},
	urldate = {2026-05-24},
	publisher = {SAGE Publications Ltd},
	author = {Schreier, Margrit},
	year = {2012},
}

\end{multicols}

\end{document}